\documentstyle[aps,prb,epsf]{revtex}
\begin{document}

\newcommand{\ignore}[1]{{\typeout {skipping things}}}

\title{Acceptor binding energies in GaN and AlN}

\author{
 Francisco Mireles and Sergio E. Ulloa
       }

\address {
 Department of Physics and Astronomy, and Condensed Matter and
 Surface Sciences Program \\ Ohio University, Athens, Ohio 45701--2979
         }

\date{Mar 12 1998}

\twocolumn

\maketitle

 \begin{abstract}

We employ effective mass theory for degenerate hole-bands to calculate
the acceptor binding energies for Be, Mg, Zn, Ca, C and Si
substitutional acceptors in GaN and AlN. The calculations are
performed through the 6$\times $6 Rashba-Sheka-Pikus and the
Luttinger-Kohn matrix Hamiltonians for wurtzite (WZ) and zincblende
(ZB) crystal phases, respectively.  An analytic representation for the
acceptor pseudopotential is used to introduce the specific nature of
the impurity atoms. The energy shift due to polaron effects is also
considered in this approach. The ionization energy estimates are in
very good agreement with those reported experimentally in WZ-GaN.  The
binding energies for ZB-GaN acceptors are all predicted to be
shallower than the corresponding impurities in the WZ phase.  The
binding energy dependence upon the crystal field splitting in WZ-GaN
is analyzed.  Ionization levels in AlN are found to have similar
`shallow' values to those in GaN, but with some important differences,
which depend on the band structure parameterizations, especially the
value of crystal field splitting used.

 \end{abstract}


\narrowtext

\section{Introduction}

Wide-band gap III-V nitrides, particularly Ga-, Al- and InN, and their
semiconductor alloys, are materials currently under intense study.
Some of their most promising applications in optoelectronics devices
are for instance the fabrication of blue/green LED's, \cite{led} laser
diodes, \cite{laser} and `solar-blind' UV photodetectors. \cite{UV}  The
performance improvements of these and related optoelectronic devices
depend strongly on the features of the intrinsic and extrinsic
impurity defects in the nitride compounds.  For example, defects and
impurities provide free carriers under suitable conditions.
Therefore, knowing the accurate position of the donor and acceptor
levels of these systems is an issue of great importance in the
understanding of optical properties and practical applications of
these nitrides.

At present, Mg and Zn are the impurity materials most widely employed
in the p-doping of GaN. The experimental thermal ionization energy
(acceptor binding energy) associated with Mg is estimated at 250
meV. \cite{Strite} The highest doping achieved reaches hole
concentrations of approximately $3\times 10^{18}$ cm$^{-3}$ at room
temperature. \cite{conc} It is also known that in order to activate
the dopants and improve the p-type conductivity, the samples must be
treated with low energy electron beam irradiation, furnace
annealing, or rapid thermal annealing after growth.\cite{leebi}
On the other hand, Zn doping seems to be inefficient because of its relatively deep ionization energy (340 meV). \cite{Strite}  Other 
dopants have been considered, but experimental problems like 
instability and/or hole compensation due to the formation 
of acceptor-H neutral complexes is still at issue. Estimates 
for the binding energies of several substitutional acceptors in GaN
have been obtained in the past, mostly through photoluminescence 
(PL) spectra.\cite{Strite} However residual impurities and defects in this material complicates the identification of these levels.
In contrast, little is known about the doping and spectrum of 
impurity levels in AlN. In fact, no conclusive results for the 
doping of AlN with sufficiently high conductivity have yet been reported.

 Apart from the question of successful p-doping in GaN and AlN using
various impurities, there are still at least two other important issues
that are under scrutiny. The first one is related to
the determination of the origin of the chemical shift observed in the
acceptor spectrum levels in GaN, apparently induced by the differences
in the cores of the various impurity atoms, and some possible lattice
relaxation around the impurity atom. \cite{Strite}  The second
question is whether acceptors with smaller binding energies ($< 230 $
meV) exist for wurtzite and zincblende GaN. The occurrence of
relatively large ionization energies for acceptors in GaN has been
attributed in part to the fact that the III-V nitrides are more ionic
than other III-V compounds (such as GaAs, GaP and InP), for which the
acceptor binding energies are an order of magnitude smaller than those
found in GaN.  It has also been suggested that the enhanced binding
energies found for some acceptors like Zn and Cd, are associated to
the relaxation of the d-electron core. \cite{Strite-1}  On the other
hand, impurities without d-electron states, such as Mg, C, and Si,
appear to induce rather shallow acceptor levels. Indeed, very
recently, Park and Chadi \cite{Park-CH} examined the stability of
acceptor centers in GaN, AlN and BN using first principles
calculations.  They concluded that the small bondlengths in III-V
nitrides inhibit large lattice strain relaxations around impurities
(mainly Be, Mg and C), giving rise to relatively shallow states for
these species.  This would suggest that a similar lack of relaxation
accompanies other substitutional impurities in these hosts, producing
relatively shallow levels, as long as there are no d-cores close to
the valence band energies.

Very recently, the formation energies and impurity levels for a few
donor and acceptor species have been studied theoretically by several
groups, \cite{Bogus,Bernardini,Neugebauer,Fiorentini,Bernard1} 
employing quantum molecular
dynamics schemes and total energy calculations in the local density
approximation of density functional theory.  In general, consistency
is found among those groups, as well as with experimental reports for
some impurity levels, such as Mg$_{\it Ga}$ acceptors 
(X$_Y$ indicates the ion X substituting in the $Y$ site).
However, this is not the case for other acceptors like C$_{\it N}$,
where discrepancies of factors of three exist among theoretical
values.  Although the calculated energy levels for these approaches
appear reliable for most cases, the impurity levels reported for some
acceptors are close to the systematic error bars introduced in the
calculations.  The delicate and complex nature of these calculations,
which require intensive computations, suggest that alternative methods
should be explored in the study of impurity levels in these systems.
There is also, no doubt, the need for new careful experiments in the
better-characterized materials now available, to clarify these
discrepancies.  The features of the acceptor states in the different
crystal phases, wurtzite (WZ) and zincblende (ZB), have not been
discussed either. In order to address these questions, we present here
a contribution towards the theoretical treatment of the impurity
levels in GaN and AlN based in the effective mass approach for
degenerate bands.

In this paper we report effective mass theory calculations of the
acceptor binding energies for various impurity atoms in GaN and AlN
for both crystal structures, WZ and ZB. Particular attention has been
paid to chemical shifts introduced by the foreign atoms. An
acceptor-pseudopotential model is used to take into account this
effect. The approach used here is based on the effective mass theory
(EMT) for degenerate bands.  Well parameterized valence band structure
calculations are used as input.  The results obtained, with no
adjustable parameters, are in very good agreement with experiments, as
we will see below.  Inevitably, the application of even a simple
hydrogen-like model of acceptor states in group III-V semiconductors
is more complicated than for idealized semiconductors with a single,
isotropic and spin-degenerate valence band. The complications are due
in part to the band warping and sixfold degeneracy or near-degeneracy
of the valence band structure close to the $\Gamma$ point (${\bf
k}=0$).  Since the perturbing potential introduced by the foreign
atoms can be seen to zeroth order as pure Coulomb-like, the problem
can be seen as a generalized hydrogenic problem, where the kinetic
energy of a hole, in the rather complicated valence band structure of
the III-V materials, is properly described by a $6 \times 6$ matrix
Hamiltonian which describes well the dispersion features of the
various hole bands.  The EMT calculations of the binding energies of
Be, Mg, Zn, Ca, and C acceptor impurities are shown to be in very good
agreement with the available experiments, and consistent in general
with other theoretical calculations employing other methods (with the
exceptions discussed above for C, for example).  The applicability of
EMT for the calculation of impurity levels with 0.2--0.4 eV binding
energies is then verified {\em post facto}, likely due to the large
bandgap in these materials, which yields negligible mixing of
conduction band states.

Additionally, we find that the binding energies for acceptors in the
ZB structures are predicted to be shallower than their counterparts in
the WZ structures, suggesting that doping of ZB material would be of
significant practical advantage.  We notice that the difference on
parameters, mainly the existence of a crystal field splitting for the
WZ nitrides, strongly affects the band mixing and correspondingly the
binding energies in the two polytypes.  It is also likely that
differences in the hole masses contribute to the calculated different
binding energies.  Although substitutional impurities do not represent
a strict test of the different band parameterizations, the subtle
interplay of the different valence bands on the resulting binding
energies provides an interesting overall consistency check of the
parameterized band structure.

The paper is organized as follows. In section II, we present the
characteristics of the generalized acceptor problem. The explicit
matrix form of the ZB and WZ valence band Hamiltonians are also given
there. The trial form of the envelope wave functions is presented
in section III. The impurity pseudopotential model is discussed
in detail in section IV. The correction due to polaron effects 
is briefly described in section V. The results and discussion are 
given in section VI, and the conclusions in section VII.

\section{Generalized shallow acceptor problem}

Substitutional impurities with one fewer valence electron than the host
atom of the pure crystal introduce well localized acceptor states lying
just above the top of the valence band structure. The theory of shallow
donor and acceptor states in semiconductors has been reviewed in detail
by Pantelides. \cite{Pantelides}  We assume, as usual, that all
acceptor levels in the semiconductor are described within the effective
mass theory for degenerate band structures by the following matrix
equation

\begin{eqnarray}
{ H}({\bf r}){\bf F}({\bf r}) & = & [{\cal H}({\bf r})+U({\bf r})
                                 {\bf 1]F}({\bf r}) \nonumber
			   \\ & = & E{\bf F}( {\bf r}) \, ,
\label{ec1}
\end{eqnarray}
 where ${ H}({\bf r})$ is the full acceptor Hamiltonian with
eigenvalues $ E$ for the acceptor states. Here, ${\cal H}({\bf r})$ is
the Hamiltonian properly constructed from crystal symmetry
considerations which entirely describes the spectrum and eigenvalues
of a hole near the valence band extremum at the $\Gamma$ point.
Symmetry invariance group theory \cite{Pikus,RSP} and {\bf k$\cdot$p}
perturbation theory for degenerate bands \cite{Lutt,Chuang} has been
used to derive the proper effective-mass Hamiltonian for strained
semiconductors depending upon the crystal structure.

The potential $U({\bf r})$ is the perturbation produced by the
acceptor-ion on the otherwise pure and periodic host crystal.  In a
simple idealized case, $U({\bf r})$ is taken to be the Coulomb
potential $U({\bf r})=e^2 / \epsilon _o|\bf r|$, where $\epsilon _o$
is the static dielectric constant of the crystal, $\epsilon
(q=0,\omega=0)$, representing a point charge in a dielectric medium.
Notice that the screening of the simple hydrogenic potential by a
dielectric function $\epsilon(\bf q)$ has been considered in the past
as an approach to consider the contribution to the acceptor spectrum
of the short range potential from the real impurity.\cite{Bernholc}
Although this model gives an insight into the specific character of
the different atomic acceptor levels, the model results in a generic
value for all the impurity defects.  This, clearly, neglects the
chemical signature of the foreign atoms in the host material (the
so-called central-cell contribution). \cite{Pantelides}  Given these
limitations, we employ instead an {\em ab initio} pseudopotential
$U_{ps}(r)$ corresponding to the difference between the bare model
potential of the impurity and the host atoms. Since the chemical
correction induced by different species is expected to be small and
because the pseudopotential used is fairly smooth and without
discontinuities, the effective mass approach is expected to yield an
appropriate description of the system.  More details on the impurity
potentials used are given in section IV.

In Eq.\ (\ref{ec1}), ${\bf 1}$ is the $6\times 6 $ unit matrix and
${\bf F}({\bf r})$ is a column vector whose $F_j({\bf r})$ elements
characterize the envelope function which modulates the Bloch functions
$\phi _j({\bf r})$ of the unperturbed crystal at the top $({\bf
k}\approx 0)$ of the valence structure. Correspondingly, the wave
functions for the shallow states are given by

\begin{equation}
 \psi ({\bf r})=\sum_{j=1}^6F_j({\bf r})\phi _j({\bf r}) \, .
 \label{ec2}
 \end{equation} 
 The trial form chosen for the envelope functions $F_j({\bf r})$ is
discussed in detail in section III. In the following subsections we
describe briefly the explicit form of the hole Hamiltonian ${\cal
H}({\bf r})$ for the two crystal polytypes (WZ and ZB), in which the
bulk GaN and AlN semiconductors grow.

\subsection{Wurtzite valence band Hamiltonian}

In order to consider the motion of a carrier at the top of the valence
band in a wurtzite semiconductor we must take into account its
six-fold rotational symmetry, which induces a crystal field
splitting. Moreover, in the case of spin-orbit interaction, the
$\Gamma _{15}$ level splits into the $\Gamma _9$ state, upper $\Gamma
_7$ level, and lower $\Gamma_7$ level, corresponding to the heavy
hole, light hole and split-off hole bands. \cite{Pikus} The
appropriate effective mass Hamiltonian that reflects those features of
the WZ GaN bulk crystal should be described thus by the
Rashba-Sheka-Pikus Hamiltonian (RSP),\cite{Pikus,RSP} as discussed
recently by Sirenko {\em et al}.\cite{Sirenko} In the vicinity of the
valence band maximum, and to second order in $k$, the six states
(including the spin index) of the RSP Hamiltonian for unstrained WZ
structures can be explicitly written in a matrix representation as
follows:

\begin{equation}
{\cal H_{WZ}}({\bf k})=\left(
\begin{array}{cccccc}
F & 0 & -H^{*} & 0 & K^{*} & 0 \\
0 & G & \Delta & -H^{*} & 0 & K^{*} \\
-H & \Delta & \lambda & 0 & I^{*} & 0 \\
0 & -H & 0 & \lambda & \Delta & I^{*} \\
K & 0 & I & \Delta & G & 0 \\
0 & K & 0 & I & 0 & F
\end{array}
\right) \, ,
\label{ec3}
\end{equation}
where
\begin{eqnarray}
 F &=& \lambda +\theta +\Delta _1+\Delta _2 \,\ \nonumber \\
 G &=& \lambda +\theta +\Delta _1-\Delta _2  \nonumber \\
 \lambda &=& A_1k_z^2+A_2k_{\perp }^2 \qquad \qquad   \nonumber  \\
 \theta &=& A_3k_z^2+A_4k_{\perp }^2  \nonumber \\
 H &=& i(A_6k_zk_{+}+A_7k_{+})   \nonumber \\
 I &=& i(A_6k_zk_{+}-A_7k_{+})  \nonumber \\
 K &=& A_5k_{+}^2 \nonumber \\
 \Delta &=& \sqrt{2}\Delta_{3}\, ,
\end{eqnarray}
\noindent
  with $k_{\perp }^2 = k_x^2+k_y^2$, and $k_{\pm }=k_x\pm ik_y$.
Here, $\Delta _1$ corresponds to the energy splitting produced by the
anisotropy of the hexagonal symmetry, $\Delta _2=\Delta^{(z)}_{so}/3$
and $\Delta _3=\Delta^{(\perp)}_{so}/3$
are the energy splittings for the $z$ and perpendicular directions
produced by the spin-orbit (SO) interaction.\cite{Sirenko}  The
$A$ constants are related to the inverse of the hole masses, in units
of $\hbar^2/2m_o$, where $m_o$ is the bare electron mass.  Notice that
when the linear terms in (\ref{ec3}) are negligible ($A_7=0 $; which is
in fact nearly the case in GaN and AlN), the RSP Hamiltonian has
complete inversion symmetry.  This symmetry allows for helpful
simplifications in dealing with the acceptor problem in the envelope
function framework, as we discuss below.

\subsection{Zincblende valence band Hamiltonian}

In the case of semiconductors with the ZB structure, the hole wave
functions characterizing the sixfold degenerate $\Gamma _{15}$ state
split, due to the effects of spin-orbit interaction, into the fourfold
degenerate $\Gamma _8$ states corresponding to the heavy and light hole
bands, and the spin-split off hole states $ \Gamma _7$. \cite{Pikus}
The Hamiltonian which takes into account all these features of the
cubic symmetry for ZB semiconductors is the well known Luttinger-Kohn
Hamiltonian (LK),\cite{Lutt} which at the valence band extremum, and to
second order in $k$, is expressed in terms of only four empirical
parameters --- the so-called Luttinger-Kohn parameters $\gamma
_1,\gamma _2$ and $\gamma _3$, and the spin-orbit splitting
$\Delta _{o}$.  Thus, the LK Hamiltonian ${\cal H_{ZB}}$ is written in
matrix form as follows

\begin{equation}
{\cal H_{ZB}({\bf k})=}\left(
\begin{array}{cccccc}
   P        &       L       &       M       &       0     &       N       &      S       \\
  L^{*}     &       Y       &       0       &       M     &       R       &   \sqrt{3}N  \\
  M^{*}     &       0       &       Y       &      -L     & \sqrt{3}N^{*} &       R      \\
  0         &      M^{*}    &     -L^{*}    &       P     &    -S^{*}     &     N^{*}    \\
  N^{*}     &      R^{*}    &    \sqrt{3}N  &      -S     &      W        &        0     \\
  S^{*}     & \sqrt{3}N^{*} &     R^{*}     &       N     &      0        &      W
 \end{array} \right) \, ,   
 \label{ec3.1}
\end{equation}
 where
 \begin{eqnarray}
 L &=& -2\sqrt{3}i\gamma _2k_zk_{-}  \nonumber \\
 M &=& \sqrt{3}\gamma _3(k_x^2-k_y^2)-2\sqrt{3}i\gamma _3k_xk_y  \nonumber \\
 N &=& \frac i{\sqrt{2}}L  \nonumber \\
 P &=& \gamma _1k^2-\gamma _2\left( 2k_z^2-k_{\perp }^2\right)    \nonumber \\
 Q &=& \gamma _1k^2+\gamma _2\left( 2k_z^2-k_{\perp }^2\right)    \nonumber \\
 R &=& -\frac{\sqrt{2}}3i(P-Q)   \nonumber \\
 S &=& -i\sqrt{2}M   \nonumber \\
 W &=& \frac 13(2P+Q)+\Delta _o   \nonumber \\
 Y &=& \frac 13(P+2Q) \, .
 \end{eqnarray}
 \noindent
 Here, $L, M, P$ and $Q$  are in units of $\hbar ^2 /2m_o$. Notice that
the higher symmetry of the ZB structure produces a much simpler ${\cal
H} (r)$ and fewer parameters than for the RSP case.  In both cases, the
operator ${\cal H}({\bf r})$ is obtained via the usual transformation $
k_\alpha \rightarrow i\frac \partial {\partial x_\alpha }$ in ${\cal
H}( {\bf k})$.

\section{Trial form for the envelope functions}

To solve the effective mass equation for degenerate bands, Eq.\
(\ref{ec1}), we use the fact that the effective mass Hamiltonian is
invariant under inversion with respect to the origin, so that the
envelope functions $F_j({\bf r})$ can be chosen to have definite
parity. Since the features of the acceptor problem are rather like
those of a hydrogenic-like problem, it has proved convenient to choose
the envelope functions basically as an expansion in spherical harmonics
and a linear combination of hydrogenic-like radial functions.  In
particular, we have chosen the following explicit form:

\begin{equation}
F_j({\bf r})=\sum_{l,m}f_l^{~j}(r)Y_{lm}
(\theta ,\phi) \, ,
\label{ec4}
\end{equation}
 summing over all $l$ even (or odd), and with radial functions for a
given hole band $j$ and angular momentum quantum number $l$ of the form
\begin{equation}
f_l^{~j}(r)=\sum_{i=1}^NA_i^{~j}r^le^{-\alpha _ir}\, .
\label{ec5}
\end{equation}
 In this work, however, we are mostly interested on the ground state
(the highest binding acceptor state), and in such state only even $l$
will contribute to the expansion --- as one would expect a ground
state with even parity.  This convenient simplification can be relaxed
straightforwardly if desired, with little effect on the results.  For
numerical convenience, we find it useful to minimize or evaluate the
acceptor binding energy choosing $\alpha _i^{\prime }s$ in the
progression $\alpha _k=\alpha _1e^{\beta (k-1)}$, such that
 $
\beta= (N - 1)^{-1} \log (\alpha_N/\alpha_1)
 $,
 and the end point conditions are chosen as $\alpha _1=1.2\times
10^{-2}\ a_o^{*^{-1}}$, and $ \alpha _{N}=3.5\times
10^2~a_o^{*^{-1}}$.  Here, $a_o^{*}= \tilde{\gamma_1}\epsilon _o a_o$ 
is the effective Bohr radius, and $\tilde{\gamma_1}$ is defined by

\begin{equation}
 \tilde{\gamma_1} =\left\{ \begin{array}{ll} -(2m_o /\hbar^2) (A_2+A_4) 
			& \mbox{ for WZ} \\ 
		\gamma_{1} & \mbox{ for ZB}
                         \end{array} \right. \, ,
\label{ec101}
\end{equation}
 such that the effective Rydberg energy is defined as $E_o^{*}= m_o e^4
/ 2\hbar^2 \tilde{\gamma _1} \epsilon _o^2 = e^2/2a_o \tilde{\gamma_1}
\epsilon_o$.  The range of $\alpha_i$ values was designed to cover a
wide spectrum of length scales.  In the limit case of $\tilde{\gamma
_1}=\epsilon _o=1$ (with N=25 and for $l=0,2$), one obtains the hydrogen spectrum,
so that for the first five states we obtain (in Rydbergs) $E_1=1.0000,$
$E_2=0.2500,$ $E_3=0.1111,$ $E_4=0.0625$ and $ E_5=0.0399$, as
expected.

\section{Impurity atom pseudopotential}

As mentioned above, a simple hydrogenic (scaled Coulomb) potential
would not yield the observed variations in the binding energy of
acceptor states for different impurity atom species.
Photoluminescence measurements show indeed important differences in
the acceptor binding energies in WZ GaN for different impurities \cite{Strite}. To study those `chemical' shifts one needs to use impurity potentials properly constructed to insure that their physical properties reflect the expected shifts.  The impurity potential here is obtained from an
analytical representation of the pseudopotential for the bare impurity
and host atoms.  The analytic form follows Lam {\it et al.},\cite{Lam}
who fit the first-principles pseudopotentials developed earlier by
Zunger {\it et al.} \cite{Zunger} in a density functional formalism.
Notice then that the acceptor potential is truly an impurity
pseudopotential, having its origin in {\it ab initio} calculations.
The pseudopotential for a bare atom can be written as \cite{Lam}

 \begin{equation}
 U_{ps}(r)=\sum_lV_{ps}^l(r) \hat{P_l}-\frac{Z_v}r,
\label{ec8}
 \end{equation}
 with
 \begin{equation}
 V_{ps}^l(r)=\frac{C_1^l}{r^2}e^{-C_2^lr}-\frac{Z_c}re^{-C_3r},
\label{ec9}
 \end{equation}
 where $V_{ps}^l(r)$ represents the atomic core pseudopotential. $
\hat{P_l}$ is the projection operator which picks out the component of
the wave function with angular momentum number $l$.  The
constants $C_1^l,$ $C_2^l$ and $C_3$ are the fitted parameters, with
$Z_c$ and $Z_v$ representing the core and valence electron charges,
respectively, as defined by Lam. \cite{Lam}  The first term in
(\ref{ec9}) represents a potential barrier which replaces the kinetic
energy of the true valence states, while the second term arises from
electrostatic screening of the nucleus by the core electrons and
exchange-correlation forces. Using these pseudopotentials, the impurity
model potential is constructed as follows. 

When the substitutional impurity atom replaces
the host atom in the crystal, the impurity potential is defined as the
difference between the impurity and host ion pseudopotentials. If
$l=0$, for instance,

\begin{equation}
U(r)=\frac {e^2}{\epsilon _o}\Delta V_{ps}^o(r)-\frac{\Delta
Z_v e^2}{\epsilon _or } \, ,
\label{ec10}
\end{equation}
 with
 \begin{equation}
 \Delta V_{ps}^o(r) = \pm (V_{ps,host}^o(r)-V_{ps,imp}^o(r)) \mbox{ for }
	Z_{host} ~^>_< Z_{imp} \, .
\label{ec11}
\end{equation}
 Here, $\Delta Z_v=Z_v^{host}-Z_v^{imp}$ ($=1$ for single acceptors),
and $\epsilon _o$ is the dielectric constant of the host
lattice. Clearly the first term in $U(r)$ corresponds to the net
potential produced by the difference between the bare core potentials
of the impurity and the host; it is the short-range part. The last
term is the long-range Coulombic potential due to the difference in
the valence charge $\Delta Z_v$. The static dielectric constant
$\epsilon _o$ is introduced here to reflect the effect of the lattice
polarizability (screening) of the host crystal.  Notice that in this
approach the net effect of the redistribution of charge near the
impurity defect and the accompanying screening of the foreign charge
at `large' distances (several lattice units) is considered fully in
the pseudopotential definition.

In a different approach, frequent in the literature, \cite{Wang} the
role of the pseudopotential is partly simulated using a $q$-dependent
screening function $\epsilon (q) ~(\rightarrow \epsilon_o {\it ~for~} q
\rightarrow 0)$ in the simple hydrogenic-style impurity potential.  We
avoid using $\epsilon (q)$ thanks to the impurity-specific
pseudopotential.  We believe our approach to be better in this problem,
as it requires no further adjustable parameters and yields the expected
chemical shifts quite accurately.

To provide a simple and independent test of the model we have
calculated the binding energies for several acceptors in the well
characterized semiconductor GaAs. The results are shown in Table I. The
theoretical binding energies are in excellent agreement with the
experimental values, with no additional parameters.

\section {Polaron correction}

We should also notice that since the nitride semiconductors (GaN and
AlN) are polar materials, one would expect that the
electron-LO phonon coupling would introduce
corrections to the bound states.  In order to obtain an estimate of
such correction, we assume that the polaron contribution to the
acceptor binding energy close to the $\Gamma$ point is diagonal in
band index. Therefore the acceptor binding energies will be enhanced
by $(1+ \alpha_F (m^*_j)/6 )E^*_{o,j}$, up to first order in the
Fr\"ohlich coupling constant $\alpha_F$ for each hole
band. This coupling constant is defined by \cite{Sak}

 \begin{equation}
 \alpha_{F}(m^*_j) = \left(
\frac{1}{\epsilon_\infty}-\frac{1}{\epsilon_o} \right) \left(
\frac{E_o}{\hbar \omega} \frac{m^*_j}{m_o} \right)^{1/2} \, ,
 \end{equation}
 where $E_{o}$ is one Rydberg, 
 $E^*_{o,j}$ is the ground state energy of the impurity acceptor
without the polaron correction, $\hbar \omega$ is the LO phonon energy,
and $m^*_j$ is the average $j$-hole effective mass.  In this way, the
contribution of each hole band to the polaron energy is taken into
account explicitly in the multiband calculation.  Let us notice that
the resulting polaron correction is relatively small (not greater than
8\%) in all cases, as shown in the tables below, despite the polar
nature of these materials.  This is presumably due to the fact that the
coupling constant associated to each hole band is relatively small
($\le 1.5$) in all cases. \cite{XWFMP97}

\section{Results and Discussion}

Since the reported values of effective mass parameters obtained by
different approaches for both the RSP and LK Hamiltonians may have
significant discrepancies, we have used  different sets of
parameterizations in order to  compare the resulting impurity states.
\cite{Kim,Suzuki,Maje,Yeo,Wang,Meney} 

For the wurtzite system (Tables II and III), we use Kim {\it et al.}
\cite{Kim} RSP parameterizations obtained by full potential linearized
muffin-tin orbital (FP-LMTO) band structure calculations, in which the
spin-orbit coupling effects were obtained via the atomic-sphere
approximation.  We have also used the Suzuki {\it et al.}
\cite{Suzuki} RSP parameters obtained by full potential linearized
augmented plane wave calculations (FLAPW); a different set reported by
Majewski {\it et al.} \cite{Maje} based on the norm-conserving
pseudopotential plane-wave (PPPW) method, and a fourth set obtained by
Yeo {\it et al.}, \cite{Yeo} who employ an empirical pseudopotential
method. Notice that differences in parameters between these two groups
are typically small, but can be substantial in some cases (such as the
value of the crystal field splitting $\Delta_1$), having important
consequences on the binding energy calculations, as we see later.

In the case of zincblende structures, the LK hole-parameters used are
those reported by Kim {\it et al.},\cite{Kim} and  Suzuki {\it et
al}.\cite{Suzuki} as mentioned above, and a third set by Wang {\it et
al.}, \cite{Wang} based on pseudopotential calculations.  These
parameters are summarized in Table IV.

We first examine our results for the acceptor levels in WZ nitrides.
We should emphasize here that the experimental values of the acceptor levels in WZ-GaN are not without controversy. Nevertheless, in order
to have a trend of the binding energies for different dopants, we compare
our theoretical calculations with the experimental values in the literature. For GaN the results are listed in Table V (theoretical binding energy values are reported here to the nearest meV, but are calculated with
much higher numerical accuracy for each set of parameters).  We note
that in general the binding energies for different impurities are in
good agreement with those values observed in experiments. For
instance, our calculations with the Suzuki {\em et al}.  \cite{Suzuki} 
parameters give rise to a binding energy for Be$_{\it Ga}$ and Mg$_{\it Ga}$
(241 and 253 meV, respectively, with the polaron correction included) which 
would seem to be in better accord with the reported
experimental value (250 meV).  Indeed, Salvador {\it et al.}
\cite{Salvador} reported recently room-temperature photoluminescence
spectra of Be-doped GaN films. They found strong features in the
390-420 nm range which were attributed to the acceptor state formed by
Be at about 250 meV above the valence band edge. Even though residual
impurities could be responsible as well for this level, no experiments have been reported to confirm either claim. Very recently,
Bernardini {\it et al.} \cite{Bernardini} using first principles
calculations, predict that Be is a shallow acceptor in GaN with a
binding energy (BE) of only 60 meV, in clear contrast with our calculations
and with the experimental data.  It is interesting to note, however,
that our BE's for Mg ($\sim$ 200-250 meV) are in satisfactory
agreement with those theoretical values obtained from first principles
calculations by Fiorentini {\it et al.} \cite{Fiorentini} ($\sim$\,
230 meV) and Neugebauer {\it et al.} \cite{Neugebauer} ($\sim$ 200
meV).

In contrast, the binding energies with Ref.\ [\ref{Suzuki}]
parameters, for Zn and C impurities, are overestimated with respect to
the experimental values (presumably due to the high value of crystal
field reported in [\ref{Suzuki}]).  In principle we should expect the
best fit precisely for these impurities since they are isocoric with
Ga and N, respectively, which would produce negligible local
relaxations and core polarization effects.  The best agreement occurs
when we use the parameters from Kim {\it et al.}, \cite{Kim}
suggesting that their parameter set is somewhat better. For example,
for Zn$_{Ga}$ in GaN, we obtain a BE of 331 meV using Kim's
parameters, which is in good agreement with the experimental value of
340 meV, and in excellent agreement with the theoretical value
reported by Bernardini {\it et al}. (330 meV). \cite{Bernard1}
Concerning the C$_{N}$ substitutional impurity in a N site, we find
that with exception of Suzuki's parameters, all the hole-band sets
give BE's (223-240 meV) comparable with the experimental value of 230
meV from Fischer {\it et al}. \cite{Fischer} Note that using Kim's
parameters gives the acceptor level just even with the experimental
value, in a nice but probably fortuitous agreement, considering the
possible sources of systematic errors.  Boguslawski {\it et al.}
\cite{Bogus} had predicted also an ionization energy for C$_{N}$ of
$\sim 200$ meV, while Fiorentini {\it et al.} \cite{Fiorentini} report
a deeper ($\sim 600$ meV) value. The formation energy for this
impurity is found to have also a substantial difference (1.4 eV)
between those authors. The relatively higher relaxation effects
predicted by Ref.\ [\ref{Bogus}] seem to play a more crucial role
here. Similar discrepancies are found between the present work and
other calculations for Ca$_{\it Ga}$ and Si$_{\it N}$. \cite{Bogus}
We found that  Ca$_{\it Ga}$ has its acceptor level ($\sim 260$ meV), 
close to that of the Mg. It is interesting to notice that 
temperature-dependent Hall measurements of Ca-doped GaN have
shown that the thermal ionization energy level of Ca ($\sim$\ 0.17 eV) is 
similar to that found in Mg ($\sim$\ 0.16 eV). \cite{Lee,Akasaki} 
This could indicate that the acceptor binding energy for Ca
is also close to that for Mg as we have indeed predicted.
Similarly, Si$_{\it N}$ was found to have a rather shallow level
in WZ-GaN at about 0.2 eV. While the donor behavior of Si 
is well known, no reliable experimental evidency of Si acceptor 
has been reported.

The collection of results discussed above indicates that the
parameterization of Kim {\em et al.}  \cite{Kim} leads to acceptor
binding energies in overall better agreement with the experiments and
other theoretical estimates.  Notice however that the differences in
binding energies in GaN with other sets of parameters are not large in
most cases, within a few percent from each other.
 
We would like now to comment on the effect of the crystal field
splitting on our calculations. Whereas recent experiments seems to
indicate that the $\Delta_1$ value is about 10 meV, 
\cite{Edwards,Gil,Gil1} the theoretical estimates are still controversial,
varying between 22-73 meV for GaN depending upon the approach used. \cite{Kim,Suzuki,Maje,Yeo} For example,  Ref. [\ref{Kim}] and [\ref{Suzuki}] had obtained $\Delta_1=$36, and 73 meV, respectively. The former authors 
attribute the large theoretical discrepancy to the use of an  
ideal-cell internal structure parameter 
$u$ in Ref.\ [\protect \ref{Suzuki}], instead of the relaxed one. 
In any case, to illustrate the effect of the binding energies upon 
the $\Delta_1$ value, we show in Fig.\ 1 their dependence on this 
parameter for Mg$_{\it Ga}$ in WZ-GaN, over a wide range.  
A rather monotonic behavior is seen in the binding
energies, as one would expect. Note that for all parameter 
sets (with exception of those in Ref.\ [\ref{Maje}]) the BE's
are consistently close for each $\Delta_1$ value. Note that using 
the experimental value of 10 meV for  
$\Delta_1$ would produce smaller binding energies, 
giving values of about 0.19 eV, regardless of the set employed.
The behavior for other dopants shows an analogous trend,
where the energy shift on the binding energy is 
nearly the difference in $\Delta_{1}$ values. This discussion
indicates that additional experimental evidence for a smaller
$\Delta_{1}$ value, and comparison with better optimized estimates,
would be of interest.

The results for AlN in the wurtzite structure are given in Table VI.\@
The first thing to notice here is that due perhaps to the large
discrepancy in $\Delta_1$ values, $-215$, $-219$, and $-58$ meV for
Kim, \cite{Kim} Majewski, \cite{Maje} and Suzuki, \cite{Suzuki}
respectively, the binding energies differ by almost a factor of two
for different parameter sets. Notice further that values of $A_5$ and
$A_6$ also differ substantially for different authors, strongly
affecting the band mixing and corresponding binding energies.  Given
the better agreement of Kim {\em et al}. parameters in WZ-GaN, we are
inclined to think that the corresponding results in WZ-AlN will be
perhaps closer to the experimental results.  Unfortunately, as we
mentioned earlier, the experimental spectrum for acceptors in AlN is
unknown at present (due to the well known difficulties in doping this
material \cite{Strite}).  Further scrutiny of the parameters reported
by these and future authors should be carried out to solve the
disagreements.  Notice that the BE of $C_N$ in WZ-AlN is found to
exceed 0.65 eV in our calculations for all three sets of parameters
(not shown in Table VI).
This value, in the limit of validity perhaps of our EMT calculations,
suggests nevertheless that such impurity will yield a somewhat deeper
level than those reported in Table VI.\@ Although substitutional
impurity calculations do not represent a strict test of the band
parameterizations, the subtle interplay of the different valence bands
on the resulting binding energies (or even excited impurity states)
provide an interesting overall consistency check.

For the ZB phase, we notice that predicted binding energies are
consistently smaller (by nearly a factor of two) than in the WZ
structure of GaN. Indeed, typical differences of roughly 100 meV are
found in the binding energies between the two phases (ZB and WZ) in
this material. This would have important consequences in electronic
uses once doping of ZB phases is stabilized. Concerning the resulting
impurity binding energies for GaN, we observe that the LK parameters
given by Ref.\ [\ref{Kim}], [\ref{Suzuki}], and [\ref{Wang}] give rise
to binding energies which are in close agreement with each other.  We
should also comment that a different set of band parameters in the ZB
phase has been given by Meney {\it et al}, using a semi-empirical
perturbative approach. \cite{Meney} However, using these parameters
result in BE's much smaller than those presented here.  This
difference, even greater in the binding energies for ZB-AlN, reflects
the more approximate nature of the parameters in Ref.\ [\ref{Meney}].
Notice that the Luttinger $\gamma$-parameters in Kim {\em et al}. are
slightly smaller than for Suzuki {\em et al}. (or equivalently,
slightly larger effective masses), which would be expected to yield
slightly larger BE's for the former set of parameters, as is clearly
seen in Table VII.

Recent PL spectra of cubic GaN by As {\it et al.}\cite{As} had claimed
as indeed we have predicted in our calculations, that acceptor BE's for
cubic-GaN may have energies shallower than these in wurtzite-GaN.
Acceptor energies of about 130 meV have been estimated by them. This in very good agreement with our calculations; as we can see in Table VII, the BE's ranges from  $\sim$\ 130 meV for Si to $\sim$\ 180 meV for Zn. This acceptor level has not been identified and it is probably produced by residual impurities. 

The smaller binding energies in ZB, with respect to impurities in the
WZ structure, is an interesting result that should be understood in
terms of the different band structure parameters.  Notice, however,
that the difference in the effective Rydberg energy for WZ and ZB GaN,
is not large at all, as seen in Tables II and IV.  Similarly, the
effective Bohr radius for both structures is nearly the same, as
illustrated in the fact that $\tilde \gamma_1$ is of the same order 
in both cases, and that the dielectric constant for both polytypes has 
been taken as $\epsilon_o=9.5$.  The polaron correction is certainly 
relatively small also, and is therefore not a possible source of the binding
energy difference in these polytypes.  However, the parameter that
apparently gives rise to these large shifts in the acceptor energies
could be identified with the in-plane heavy hole mass, which is indeed
larger in wurtzite than in zincblende for both GaN and AlN, and hence
produces larger binding energies.  In order to verify the effect of
the different effective masses in the two polytypes, we have
calculated the acceptor levels for WZ-GaN using the quasicubic sets of
parameters of Ref.\ [\ref{Kim}], with the same crystal field splitting
than the obtained for the non-quasicubic set. It turns out that the
binding energies are smaller correspondingly, which confirms our
assumption.  One should also mention that just as seen in Fig.\ 1, a
vanishingly small $\Delta_1$ (as is the case in ZB) would produce an
even smaller binding energy for a given impurity. [This would also
explain the agreement among the three sets of parameters, since
$\Delta_1$ differences are the most significant for different
authors.]  We then conclude that it is in fact a combination of the
crystal field splitting and slightly larger hole masses that produce
larger binding energies in WZ than in the ZB structure.  An
interesting and important effect of the different lattice and band
structure.

\section{Conclusion}

We have carried out calculations for the shallow acceptor energies
associated to different substitutional impurity atoms in GaN and AlN
hosts.  The calculations were performed within the effective mass
theory, taking into consideration the appropriate valence band
Hamiltonian symmetries for the WZ and ZB polytypes, and using the full
$6\times 6$ acceptor Hamiltonian and included the actual spin-orbit
energy splitting.  In addition, the impurity pseudopotential and the
electron-phonon (polaron) correction has been explicitly considered.
These more realistic treatment allows us to compare directly with the
observed data and verify that our calculation produces the appropriate
`chemical shifts'.  Indeed, our calculations of the acceptor binding
energies are in quite good agreement with PL experiments, as the 
introduction of the impurity 
pseudopotential seems to be an excellent
model to describe the chemical shifts associated with each impurity
atom.  It is interesting that the good fits were found without any
adjustable parameters in the calculation, once the contribution due to
the electron-phonon polar interaction was included. We find that small
differences in the hole
effective mass parameters could lead to relatively large discrepancies in the
binding energies.  Our overall evaluation of parameters suggests that
the better BE values are obtained with those in Ref.\ [\ref{Kim}].
Correspondingly, we refer the reader to the first line in each impurity
case in Tables V, VI, and VII, for what we consider the best BE
estimates, within a few percent error.  Further refinement of
experimental values would be desirable to set narrower constraints on
the theoretical values.  We also find that the binding energies for
acceptors in the ZB structures are much shallower than their
counterparts in the WZ structures, suggesting perhaps much more
efficient carrier doping in those systems (yet to be observed
experimentally).

Finally, we should mention that preliminary studies of the strain
effects on the acceptor binding energies show an increase as the strain
increases, although with a much stronger dependence than in other III-V
materials.  A complete report of these studies will be presented
elsewhere.

\acknowledgments

We thank K. Kim, W.R.L. Lambrecht and B. Segall for kindly
communicating unpublished results to us, and for very helpful
discussions. This work was supported in part by grants ONR-URISP
N00014-96-1-0782, DURIP N00014-97-1-0315, and from CONACyT-M\'{e}xico.

\clearpage

\begin{table}[h]
\begin{center}
\begin{tabular}{lcccc}
                 &           &              &              &\\
                 &   C       &      Mg      &     Zn       &\\
                 &           &              &              &\\
\hline
                 &           &              &              &\\
    Exp.         &   27.0    &    28.7      &    30.6      &\\ 
                 &           &              &              &\\
    Theor.       &   27.4    &    27.7      &    28.3      &\\
                 &           &              &              &\\
\end{tabular}
\end{center}
\caption{Comparison between the experimental and the EMT-pseudopotential
model  of the acceptor binding energy for various
impurities species in GaAs ($\epsilon_o = 12.4$). The energies are 
in meV. The experimental values and the band parameters are taken from Ref.\ 
[\protect \ref{Madelung}].}
\end{table}

\begin{table}[h]
\begin{center}
\begin{tabular}{lccccccc}
& \multicolumn{4}{c}{WZ-GaN}\\
\hline
                   &          &          &            &          &\\
Ref.  & [\protect \ref{Kim}]      & [\protect \ref{Suzuki}] & [\protect \ref{Maje}]     & [\protect \ref{Yeo}]     & \\
                   &          &          &            &          &\\          
\hline
 $A_1$             & --6.4     & --6.27    &    --6.4    & --7.24    &\\
 $A_2$             & --0.5     & --0.96    &    --0.8    & --0.51    &\\
 $A_3$             &   5.9     &   5.70    &      5.93   &   6.73    &\\
 $A_4$             & --2.55    & --2.84    &    --1.96   & --3.36    &\\
 $A_5$             & --2.56    & --3.18    &    --2.32   &   3.35    &\\
 $A_6$             & --3.06    & --4.96    &    --3.02   & --4.72    &\\
 $\Delta_1$        &  36       &   73      &      24     &    22     &\\
 $\Delta_2$        &  5.0      &  5.4      &     5.4     & 11/3      &\\
 $\Delta_3$        &  5.9      &  5.4      &     6.8     & 11/3      &\\
\hline
                   &           &           &             &           &\\
$\tilde \gamma_1$  &  2.91     &  3.80     &     2.76    &   3.87    &\\
 $E^{*}_o$         &  51.8     &  39.7     &     54.6    &   39.0    &\\
\end{tabular}
\end{center}

\caption{The Rashba-Sheka-Pikus valence band parameters for wurtzite 
GaN. The hole parameters $A_i$
are in units of $\hbar^2/2m_o$, while $\tilde\gamma_1$ is dimensionless;  
$\Delta_i$ values represent the energy splittings in meV;
$E^*_o$ is the effective Rydberg energy in meV. 
We use $\epsilon_o = 9.5$ as the dielectric constant in GaN. 
Signs of $A_5$ and $A_6$ parameters of  Ref.\ 
[\protect \ref{Kim}] have been changed to be consistent with those in the
definition of the usual RSP Hamiltonian.} 
\end{table}

\begin{table}[h]
\begin{center}
\begin{tabular}{lcccc}
& \multicolumn{3}{c}{WZ-AlN} \\
\hline
                   &          &           &           &\\
Ref.  & [\protect \ref{Kim}]      & [\protect \ref{Suzuki}] & [\protect \ref{Maje}]     & \\ 
                   &          &           &           &\\          \hline
 $A_1$             & --3.86   & --4.06    &   --3.82  &\\
 $A_2$             & --0.25   & --0.26    &   --0.22  &\\
 $A_3$             &   3.58   &   3.78    &     3.54  &\\
 $A_4$             & --1.32   & --1.86    &   --1.16  &\\
 $A_5$             & --1.47   & --2.02    &   --1.33  &\\
 $A_6$             & --1.64   & --3.04    &     1.25  &\\
 $\Delta_1$        & --215    & --58      &   --219   &\\
 $\Delta_2$        &  6.8     &    6.8    &     6.6   &\\
 $\Delta_3$        &  5.7     &    6.8    &     6.7   &\\
\hline
                   &          &           &           &\\
 $\tilde \gamma_1$ &  1.57    &   2.12    &     1.38  &\\
 $E^{*}_o$         &  119.9   &   88.8    &   136.5   &\\
\end{tabular}
\end{center}
\caption{The Rashba-Sheka-Pikus valence band parameters for wurtzite
AlN. We use $\epsilon_o$ = 8.5 as the dielectric constant in AlN.
Parameters have the same units as indicated in Table II. Notice the
enormous discrepancy in the crystal field splitting $\Delta_{1}$ between 
Ref.\ [\protect \ref{Kim}], [\protect \ref{Maje}], and
[\protect \ref{Suzuki}]. } 
\end{table}

\begin{table}[h]
\begin{center}
\begin{tabular}{lcccccccc}
& \multicolumn{3}{c}{ZB-GaN}
& \multicolumn{2}{c}{ZB-AlN} \\
\hline
                &           &        &       &          &       & \\
Ref.  & [\protect \ref{Kim}]  & [\protect \ref{Suzuki}] & [\protect \ref{Wang}] & [\protect \ref{Kim}]  & [\protect \ref{Suzuki}] &\\
                &           &        &       &          &       &\\
\hline
$\gamma_1$      &  2.46     &  2.70  & 2.94  &    1.40  &  1.50 & \\
$\gamma_2$      &  0.65     &  0.76  & 0.89  &    0.35  &  0.39 & \\
$\gamma_3$      &  0.98     &  1.07  & 1.25  &    0.59  &  0.62 & \\
$\Delta_o$      &  19       &  20    &  17   &    19    &   20  & \\
\hline
                &           &        &       &          &       &\\
 $E^{*}_o$      &  51.3     &  55.8  &  49.3 &    134.5 & 125.5 & \\
 \end{tabular}
 \end{center}

\caption{The Luttinger-Kohn valence band parameters for zincblende GaN
and AlN.  Here the dimensionless $\gamma_i$ are the hole band
parameters; $\Delta_o$ is the energy splitting due to spin-orbit
interaction at the $\Gamma$ point, and $E^{*}_o$ are given in meV.}

\end{table}
\clearpage

\begin{table}
\begin{center}
\begin{tabular}{lcccccc}
& \multicolumn{3}{c}{WZ-GaN}
\\ \hline & & & & &\\ 
Impurity$_{\it site}$ & $E^{*}_ b$  &  $E^{**}_ b$ & $E_ b (Exp.)$ &\\ 
& & & & &\\ 
\hline
       Be$_{Ga}$    &   193     &      204     &             &\\
                    &   233     &      241     &  250$^a$    &\\
                    &   195     &      208     &             &\\
                    &   185     &      193     &             &\\
\hline
       Mg$_{Ga}$    &   204     &      215     &             &\\
                    &   245     &      253     &  250$^b$    &\\
                    &   208     &      221     &             &\\
                    &   197     &      204     &             &\\
\hline
       Zn$_{Ga}$    &   321     &      331     &            &\\
                    &   411     &      419     &  340$^b$   &\\
                    &   394     &      406     &            &\\
                    &   352     &      360     &            &\\
\hline
       Ca$_{Ga}$    &   248     &      259     &            &\\
                    &   297     &      305     &            &\\
                    &   264     &      276     &            &\\
                    &   247     &      255     &            &\\
\hline
       C$_{N}$      &   220     &      230     &            &\\
                    &   264     &      272     &  230$^c$   &\\
                    &   228     &      240     &            &\\
                    &   214     &      223     &            &\\
\hline
       Si$_{N}$     &   192     &      203     &            &\\
                    &   231     &      239     &            &\\
                    &   193     &      205     &            &\\
                    &   183     &      191     &            &\\ 
\end{tabular} 
\end{center} 
\small{
$^a$Reference [31]

$^b$Reference [4]

$^c$Reference [32]
}
\vspace{0.1in}
 \caption{Comparison between the calculated acceptor binding energies
and experimental values for different substitutional impurities in wurtzite
GaN. $E^{**}_ b$ and $E^{*}_ b$ denotes the estimated binding energies
with and without the polaron correction. All energies are in meV. The
binding energies are obtained with the band-parameters from Ref.\
[\protect \ref{Kim}], [\protect \ref{Suzuki}], [\protect
\ref{Maje}], and [\protect \ref{Yeo}], respectively, arranged in
descending order for each impurity.} 
\end{table}

\begin{table}[h]
\begin{center}
\begin{tabular}{lcccc}
\multicolumn{5}{c}{WZ-AlN} \\
\hline \\
 Impurity$_{\it site}$&
$E^{*}_ b$ & $E^{**}_ b$ & \\ & & &\\ 
\hline
\hline

       Be$_{Al}$    &   223     &      262     &\\
                    &   446     &      472     &\\
                    &   283     &      253     &\\
\hline
       Mg$_{Al}$    &   465     &      514     &\\
                    &   758     &      795     &\\
                    &   721     &      789     &\\
\hline
       Zn$_{Al}$    &   219     &      255     &\\
                    &   438     &      464     &\\
                    &   273     &      343     &\\
\hline
       Ca$_{Al}$    &   204     &      240     &\\
                    &   376     &      402     &\\
                    &   203     &      273     &\\
\hline
       Si$_{N}$     &   214     &      250     &\\
                    &   415     &      441     &\\
                    &   245     &      315     &\\
\end{tabular}
\end{center}
 \caption{Calculated acceptor binding energies for different impurities
in wurtzite AlN.  Binding energies are ordered in descending order for
parameters from Ref.\ [\protect \ref{Kim}], [\protect \ref{Suzuki}], and
[\protect \ref{Maje}], respectively.  The large discrepancy in the
calculated values is mostly due to the important differences in the
crystal field splitting used.} 
 \end{table} 

\begin{table}[h]
\begin{center}
\begin{tabular}{lccccccc}
& \multicolumn{2}{c}{ZB-GaN} & \multicolumn{2}{c}{ZB-AlN} \\
\hline
                   &             &                &             &
               &\\
   Impurity        &  $E^{*}_ b$ &  $E^{**}_ b$   &  $E^{*}_ b$ &    $E^{**}_
b$ &\\
                   &             &                &             &
               &\\
\hline
       Be          &  124      &     133       &   265     &   292        &\\
                   &  117      &     125       &   248     &   273        &\\
                   &  126      &     133       &           &              &\\
\hline
       Mg          &  130      &     139       &   333     &   360        &\\
                   &  123      &     130       &   305     &   330        &\\
                   &  133      &     140       &           &              &\\
\hline
       Zn          &  170      &    178        &   261     &   288        &\\
                   &  155      &    162        &   245     &   269        &\\
                   &  177      &    184        &           &              &\\
\hline
       Ca          &   153     &    162        &   242     &   268        &\\
                   &   143     &    151        &   227     &   252        &\\
                   &   157     &    164        &           &              &\\
\hline
       C           &   138     &    147        &   353     &   380        &\\
                   &   130     &    138        &   320     &   345        &\\
                   &   141     &    148        &           &              &\\
\hline
       Si          &  123      &     132       &   255     &   281        &\\
                   &  117      &     125       &   239     &   264        &\\
                   &  125      &     132       &           &              &\\
\end{tabular}
\end{center}
 \caption{Acceptor states for zincblende GaN and AlN.  The three values
shown for impurities in GaN correspond to those calculated with the
parameterizations given by  Ref.\ [\protect \ref{Kim}], [\protect
\ref{Suzuki}], and [\protect \ref{Wang}], respectively.  The two values
for AlN correspond to Ref.\ [\protect \ref{Kim}], and [\protect
\ref{Suzuki}].}
 \end{table}

\begin{figure}[ht] 
\epsfysize=3.in
\epsfxsize=3.5in
\epsfbox{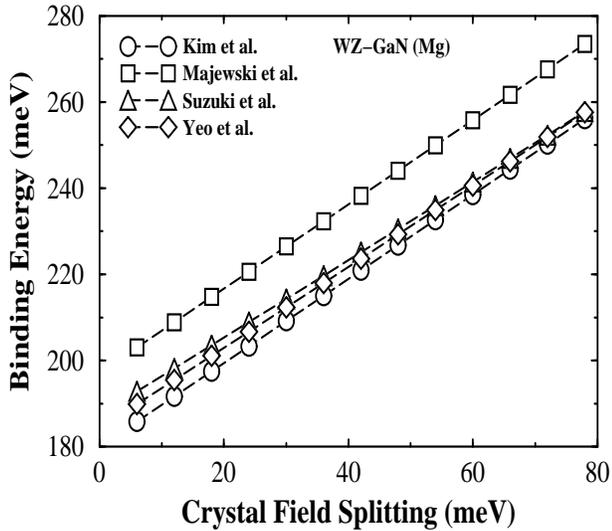}  
 \caption{{ Binding energies vs.\ crystal field splitting
$\Delta_1$, in the Mg$_{Ga}$ WZ-GaN system for different
parameterizations.  Different symbols correspond to binding energy
values obtained from the effective mass parameters, $\circ$
for Ref. [\protect \ref{Kim}], $\bigtriangleup$ for Ref. [\protect
\ref{Suzuki}], $\Box$ for Ref.\ [\protect \ref{Maje}], and $\Diamond$ for
Ref.\  [\protect \ref{Yeo}], respectively.}}
 \end{figure} 

\end{document}